\documentclass[12pt]{article}
\usepackage[dvips]{color,graphicx}
\DeclareGraphicsExtensions{.eps}

\newcommand{\ap}{\ensuremath{\alpha'}} 

\def\p{\partial}

\newcommand{\tr}{\mathop{\rm Tr}}

\newcommand{\cH}{{\mathcal{H}}}
\newcommand{\cL}{\mathcal{L}}

\newcommand{\cN}{{\mathcal{N}}}

\newcommand{\cP}{{\mathcal{P}}}

\newcommand{\cT}{{\mathcal{T}}}
\newcommand{\cU}{{\mathcal{U}}}

\newcommand{\bS}{{\mathbf{S}}}

\newcommand{\bcH}{{\bar{\mathcal{H}}}}
\newcommand{\bcL}{\bar{\mathcal{L}}}
\newcommand{\bcP}{\bar{\mathcal{P}}}
\newcommand{\bt}{\bar{t}}
\newcommand{\bx}{\bar{x}}

\title{\bf The Energy of a Moving Quark-Antiquark \\ Pair in an $\cN=4$ SYM Plasma}
\author{Mariano Chernicoff\footnote{e-mail: mariano@nucleares.unam.mx},
J.~Antonio Garc\'{\i}a\footnote{e-mail: garcia@nucleares.unam.mx}
~and Alberto G\"uijosa\footnote{e-mail: alberto@nucleares.unam.mx}
\\{\small Departamento de F\'{\i}sica de Altas Energ\'{\i}as,
Instituto de Ciencias Nucleares}\\ {\small Universidad Nacional
Aut\'onoma de M\'exico}\\
{\small Apdo. Postal 70-543, M\'exico D.F. 04510}}
\date{}

\begin{document}
\maketitle
\begin{abstract}
We make use of the AdS/CFT correspondence to determine the energy
of an external quark-antiquark pair that moves through
strongly-coupled thermal $\cN=4$ super-Yang-Mills plasma, both in
the rest frame of the plasma and in the rest frame of the pair. It
is found that the pair feels no drag force, has an energy that
reproduces the expected $1/L$ (or $\gamma/L$) behavior at small
quark-antiquark separations, and becomes unbound beyond a certain
screening length whose velocity-dependence we determine. We
discuss the relation between the high-velocity limit of our
results and the lightlike Wilson loop proposed recently as a
definition of the jet-quenching parameter.
\end{abstract}

\section{Introduction and Summary}

In the past few months, the use of the AdS/CFT correspondence
\cite{malda,gkpw,magoo} to study energy loss in finite-temperature
strongly-coupled gauge theories has attracted significant
attention, partly because it is hoped that this line of research
could eventually make contact with experimental data from RHIC
\cite{rhic} and ALICE \cite{alice}.

The drag force experienced by a heavy quark that moves through a
thermal $\cN=4$ super-Yang-Mills (SYM) plasma was determined in
\cite{hkkky,gubser}, using its dual description as a  string that
moves on an AdS-Schwarzschild background. The same information was
obtained independently in \cite{ct} through a different method,
based on an analysis of small string fluctuations (a similar
calculation was performed in \cite{hkkky}). Generalizations of the
first calculation may be found in \cite{herzog,cacg1,cacg2}. A
comparison with the corresponding weakly-coupled result was
carried out in \cite{cv}. The connection with magnetic confinement
was explored in \cite{sin2}. The directionality of the coherent
wake left by the moving quark on the gluonic fields was studied in
\cite{gubseretal1,gxz,gubseretal2}, using the methods of
\cite{dkk,cg}.

An independent approach has aimed at determining the jet-quenching
parameter $\hat{q}$ that in phenomenological models of energy loss
through medium-induced radiation is meant to codify the average
squared transverse momentum transferred to the quark by the medium
(for reviews see \cite{baier,kw}). Based on the fact that certain
approximate calculations in these models relate $\hat{q}$ to a
lightlike Wilson loop in the adjoint representation (see \cite{kw}
and references therein), the authors of \cite{liu} suggested that
this Wilson loop could be taken to provide a non-perturbative
definition of the jet-quenching parameter. Using the simple
large-$N$ relation between adjoint and fundamental loops, and the
AdS/CFT recipe for the latter\footnote{An AdS/CFT prescription for
directly computing certain Wilson loops in an arbitrary
representation of the gauge group was given recently in
\cite{gomis}.} \cite{maldawilson}, they then proceeded to compute
this parameter for $\cN=4$ SYM. Their calculation has been
generalized in various directions in
\cite{buchel,vp,cacg2,lin,sfetsos,edelstein}. Previous related
work may be found in \cite{sin1,shuryak}.

Just like the drag force determination in \cite{hkkky,gubser}, the
calculation of $\hat{q}$ in \cite{liu} focuses on a string that
moves on an AdS-Schwarzschild background. The difference is that
 the string considered in \cite{hkkky,gubser} has a single
endpoint on the boundary, representing the moving external quark,
whereas the string studied in \cite{liu} has both of its endpoints
on the boundary, representing an external quark-antiquark pair
that traces out the required lightlike Wilson loop.

In this paper we perform a natural generalization of the above
calculations, using the AdS/CFT correspondence to determine the
energy of a quark-antiquark pair that moves with velocity $v$
through a strongly-coupled thermal $\cN=4$ SYM plasma. This
problem had been previously studied in the case $v=0$, where the
pair is static with respect to the plasma. As expected, the
quark-antiquark potential was found to be insensitive to the
plasma at small distances, and to display screening behavior
beyond a certain length \cite{theisen,brandhuber}. Analyzing the
manner in which these features are modified by the motion of the
pair through the plasma is an interesting question in its own
right, both from the theoretical and the phenomenological
perspectives. Our analysis is additionally motivated by the
current discussion on energy loss: the moving quark-antiquark pair
serves as a color-neutral probe of the plasma that stands in
useful contrast with the solitary quark considered in
\cite{hkkky,gubser}, and moreover,  in the $v\to 1$ limit it would
be expected to make contact with the system studied in \cite{liu}.

Our presentation is organized as follows. In
Section~\ref{stringsec} we briefly review the salient points of
the drag force calculation of \cite{hkkky,gubser}, and then set up
and study the analogous problem for the string on
AdS-Schwarzschild that has both of its endpoints on the boundary,
satisfying the boundary conditions (\ref{bc}). Below this equation
and again after (\ref{rminfull}) we find that this string feels no
drag force, discuss the physical reasons for this result and point
out that there exist configurations with the same boundary
conditions but different initial conditions where the string
\emph{does} experience a drag force. Although framed in the
specific context of the background dual to $\cN=4$ SYM, the
essence of our arguments is more general and applies to other
backgrounds. At the end of Section~\ref{stringsec} we derive the
basic equations (\ref{yprime})-(\ref{ebar}) that determine the
shape and energy of the string in the background at hand. We work
first in the frame where the plasma is at rest, and discuss a
subtlety in defining the energy (and momentum) of the disconnected
strings dual to an unbound quark and antiquark. We then
Lorentz-transform to the frame where the $q\bar{q}$ pair is at
rest, where the energy can be defined in the standard
straightforward manner. We note that even though the string is by
definition static in this frame, it still carries momentum, which
as we explain below (\ref{energia2}) codifies information about
the momentum density of the gluonic field configuration set up by
the quarks in the dual gauge theory. Because of this non-vanishing
momentum, the energy $E$ in the plasma rest frame is not simply
proportional to the energy $\bar{E}$ in the pair rest frame. The
relation between the two is given in (\ref{energias}).

Section~\ref{qqbarsec} contains our main results, in SYM language.
We open with a discussion on the gauge-theoretic interpretation
for the result that the quark-antiquark pair feels no drag force
as it ploughs through the $\cN=4$ SYM plasma. After that we carry
out the numerical integrations needed to determine the energy of
the $q\bar{q}$ pair as a function of the separation $L$ and
velocity $v$. The results are shown in
Figs.~\ref{elow},\ref{ehigh}. The energy reduces to the expected
Coulombic behavior (\ref{Lchica}) for small separations, and then
rises above this behavior due to the effects of the plasma, up to
a screening length $L_*(v)$ beyond which the quark and antiquark
become unbound. The velocity-dependence of this screening length
is shown in Fig.~\ref{lv}; we find it to be well-approximated by
(\ref{lstar}). For velocities $v>0.447$ we find a gap in energy
between the bound and unbound $q\bar{q}$ configurations, whose
physical significance remains unclear to us.

In the closing pages of Section~\ref{qqbarsec} we discuss at
length the relation between the $v\to 1$ limit of our results and
the lightlike Wilson loop proposed in \cite{liu} as a definition
of the jet-quenching parameter $\hat{q}$. The main lesson is that
the AdS/CFT result of \cite{liu} cannot be obtained as a smooth
limit of standard Wilson loops (\ref{bc}) with $v\to 1$ from
below. We suggest that it should be regarded instead as arising
from an approach to $v=1$ from above. Finally, we note in
(\ref{q}) that, despite the fact that one cannot continuously
interpolate between the spacelike worldsheet considered in
\cite{liu} and the timelike worldsheets studied in the present
paper, the $E\propto L^2$ dependence that is central to the
definition of the parameter $\hat{q}$ in \cite{liu} is in fact
available in the $v\to 1$ limit of a subset of the configurations
analyzed here. By analogy with \cite{liu}, one can then define a
parameter $\mathcal{K}$ that, at least in this specific example,
captures exactly the same information as (and is in close
numerical agreement with) $\hat{q}$.

In the course of our investigation two related papers were posted
on the arXiv. While our work was in progress, the paper
\cite{sonnenschein} appeared, whose Section 4.2 discusses drag
effects for mesons with spin in a certain confining
non-supersymmetric gauge theory, arriving at conclusions which
coincide with those of our Section \ref{stringsec}. While the
first version of our paper was in preparation, the work
\cite{liu2} appeared, which analyzes exactly the same
quark-antiquark system as we do, focusing on the
velocity-dependence of the screening length (for an arbitrary
angle $\theta$ between the direction of motion and the
quark-antiquark separation $L$), which we determine (for
$\theta=\pi/2$) in our Section \ref{qqbarsec}. Their numerical
results are in complete agreement with ours, but as discussed
above (\ref{lstar}), their definition of the screening length
differs from ours for velocities $v<0.447$. We should also note
two additional developments that took place after the first
version of this paper had appeared: first, the addition of a plot
to \cite{sonnenschein} (Fig. 16), showing a meson size that scales
with velocity in a manner compatible with the results for the
screening length obtained in \cite{liu2} and the present paper;
second, the appearance of the work \cite{elena}, which generalizes
the screening length calculation to a large class of backgrounds,
arriving at an analytic determination of the velocity-dependence
in the ultra-relativistic regime.

\section{Nambu-Goto String in AdS-Schwarzschild}
\label{stringsec}

As mentioned in the Introduction, the computation in
\cite{hkkky,gubser} of the drag force felt by an external quark
moving through $\cN=4$ SYM plasma focuses on a string that extends
all the way from the boundary to the horizon of an
AdS-Schwarzschild geometry, i.e., from $r\to\infty$ to $r=r_H$,
with $r$ an appropriate radial coordinate that we will henceforth
also use as spatial worldsheet coordinate. We will in addition
employ the boundary time $t=x^0$ as worldsheet time, so altogether
we work in the static gauge
\begin{equation}\label{staticgauge}
\sigma=r,\quad \tau=t~.
\end{equation}
The dynamics of the string are described by the Nambu-Goto action
\begin{equation} \label{ng}
S=-\frac{1}{2\pi{\alpha}'}\int{d\tau
d\sigma}\,\sqrt{-\det{g_{\alpha\beta}}} \equiv
\frac{1}{2\pi{\alpha}'}\int{d\tau d\sigma}\,\cL~,
\end{equation}
where $G_{\mu\nu}$ is the  the spacetime metric and
$g_{\alpha\beta}\equiv
G_{\mu\nu}\partial_{\alpha}X^{\mu}\partial_{\beta}X^{\nu}$ the
induced worldsheet metric. The force that a given segment of the
string exerts along spatial direction $i$ on the neighboring
segment was expressed in \cite{gubser} as
\begin{equation} \label{forceg}
F_{i}=\frac{1}{2\pi{\alpha}'}\sqrt{-g}P^r_{i}~,
\end{equation}
with
\begin{equation} \label{smomentum}
P^{\alpha}_{\mu}=-g^{\alpha\beta}\p_{\beta}X_{\mu}
\end{equation}
the worldsheet current associated with spacetime momentum, whereas
in \cite{hkkky} it was formulated as
\begin{equation} \label{forceh}
F_{i}=\frac{1}{2\pi{\alpha}'}\Pi^r_i~,
\end{equation}
with
\begin{equation} \label{cmomentum}
\Pi^{\alpha}_{\mu}={\p\cL\over\p(\p_{\alpha}X^{\mu})}
\end{equation}
the canonical momentum densities conjugate to $X^{\mu}$. By
explicitly inverting the $2\times 2$ matrix $g_{\alpha\beta}$, one
can easily verify that
\begin{equation} \label{momenta}
\sqrt{-g}P^{\alpha}_{\mu}=\Pi^{\alpha}_{\mu}~,
\end{equation}
and so the expressions (\ref{forceg}) and (\ref{forceh}) coincide.
Clearly the latter is simpler to use in explicit computations.

The crucial point in the calculation of \cite{hkkky,gubser} is the
observation that, if this string is assumed to travel at constant
velocity $v\neq 0$ along a direction $x=x^1$ parallel to the
boundary, then there is a certain velocity-dependent value of the
radial coordinate,
\begin{equation} \label{rv}
r_v={r_H\over(1-v^2)^{1/4}}~,
\end{equation}
below which the embedding function
\begin{equation}\label{x}
X(r,t)=vt+\xi(r)
\end{equation}
 for the string would become imaginary.
 The only way to avoid this is to let the
string trail behind its boundary endpoint following a specific
profile $\xi(r)\neq$constant, which translates into a specific
value for the drag force $F_x$ exerted on the endpoint. In short,
the non-zero value of the drag force is set uniquely by the
condition that the string crosses the critical radius $r_v$.

In this section we are interested in exploring how these
considerations generalize to a moving string that has \emph{both}
of its endpoints on the boundary, and is therefore dual not to a
single quark but to a quark-antiquark pair in the gauge theory.

Since the string now extends first away from and then back to the
boundary, the static gauge choice $\sigma=r$ leads of course to a
double-valued parametrization, but this poses no problem other
than the need to check by hand that the two halves of the string
join together smoothly (which ensures that the action is
extremized not just piecewise, but over the entire worldsheet). To
describe a moving quark-antiquark pair, both of the string
endpoints are taken to travel with the same velocity $v$ in the
$x$ direction, and to be separated by a constant distance $L$
along a certain boundary direction $y=x^2$. In other words, with a
convenient choice of origin, the embedding functions (\ref{x}) and
\begin{equation}\label{y}
Y(r,t)=Y(r)
\end{equation}
 satisfy the boundary conditions
\begin{equation}\label{bc}
X(\infty,t)=~vt,\qquad Y(\infty)=\mp {L\over 2},
\end{equation}
where the upper (lower) signs refer to the left (right) half of
the string. For concreteness, we specialize immediately to the
case where $y$ is perpendicular to the direction of motion $x$,
which among other reasons is of particular interest in view of the
connection with \cite{liu}. The string starts at $r\to\infty$ and
extends down to a minimal radius $r_{min}$, which by symmetry is
such that $Y(r_{min})=0$ and $Y'(r_{min})=\infty$ (which is of
course the condition that the projections of the two halves of the
string onto the $r-y$ plane can be glued together smoothly).

\begin{figure}[htb]
\setlength{\unitlength}{1cm}
\includegraphics[width=15cm,height=8cm]{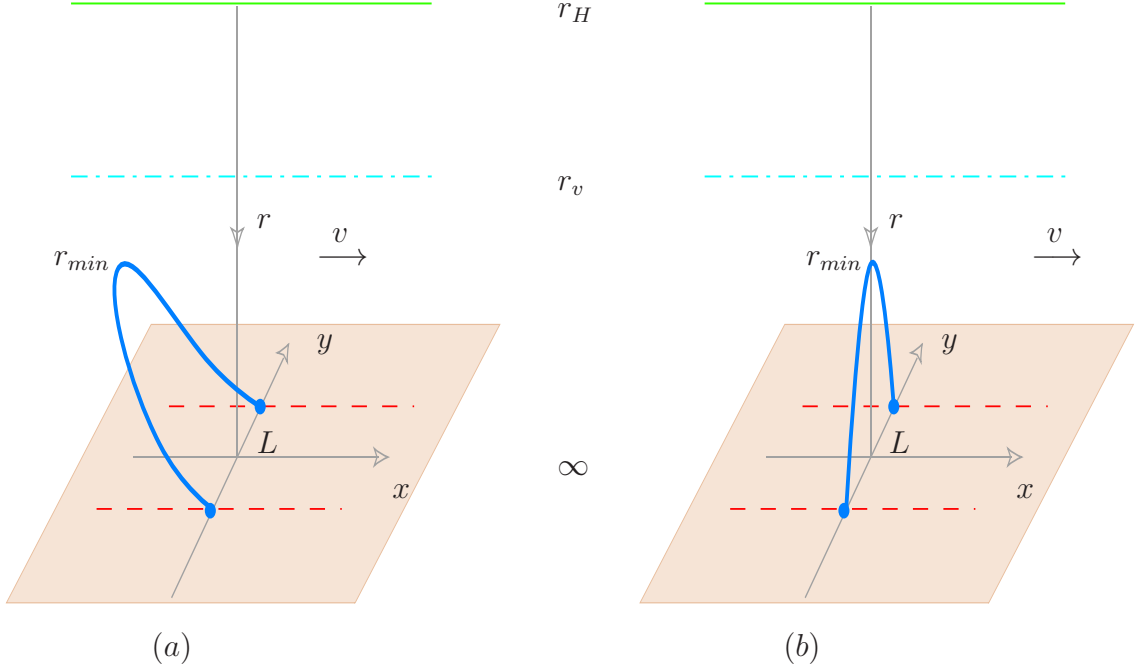}
 \begin{picture}(0,0)
\put(-13.2,-.7){$(a)$} \put(-14.5,4.5){$r_{min}$}
\put(-11,3.4){$y$}
   \put(-10,1.4){$x$}
   \put(-11.8,5){$r$}
  \put(-7.8,5.5){$r_v$}
  \put(-7.8,7.8){$r_H$}
  \put(-11.8,2){$L$}
\put(-11,4.5){$\longrightarrow$} \put(-10.8,4.8){$v$}
\put(-4.8,-.7){$(b)$} \put(-2.7,3.4){$y$}
   \put(-1.7,1.4){$x$}
   \put(-3.4,5){$r$}
 \put(-4.5,4.5){$r_{min}$}
   \put(-3.4,2){$L$}
\put(-1.5,4.5){$\longrightarrow$} \put(-1.3,4.8){$v$}
\put(-7.8,1.7){$\infty$}
\end{picture}
\vskip.5cm \caption{Sketch of the string dual to a moving
quark-antiquark pair. The radial coordinate runs downward, so the
horizon at $r=r_H$ is shown at the top and the boundary at
$r\to\infty$ is represented by the plane at the bottom. The
 dash-dot line  at $r=r_v$ marks a velocity-dependent radius beyond
 which the string cannot penetrate.
 As the string moves to the right, its
endpoints (dual to the external quark and antiquark) trace out the
dotted trajectories. Its shape
 codifies information on the configuration of the SYM color fields.
 (a) One might expect the string to lean backward as
a result of the motion. This turns out to be possible only if the
string has a nontrivial time-dependence. (b) The lowest-energy
configuration for the moving string is in fact upright, similar to
the one obtained in the static case. See text for further
discussion.} \label{string}
\end{figure}

The main question is whether such a string also trails behind its
endpoints and exerts a drag force on them, as would seem
compulsory if the string crosses $r_v$, and might appear natural
more generally on physical grounds. The shape of the string would
then be similar to that shown in Fig.~\ref{string}a. It is easy to
see, however, that this cannot happen. The reason is that, on the
one hand, the force along the direction of motion must vanish at
the midpoint ($r=r_{min}$), because by symmetry the string at that
point must be perpendicular to the $x$-axis, and on the other
hand, the force must be constant along the string, because each of
its segments moves at constant velocity. We conclude then that
$F_x=0$ everywhere, which implies that the string remains upright,
as in Fig.~\ref{string}b.

Since we initially envisioned the string as being pulled along
from its endpoints, it might  seem somewhat counterintuitive that
it does not lean backward. To clarify this point, it is worth
noting that the $r$-independence of $F_x\propto\Pi^r_x$ is a
consequence of the equation of motion for $X$ (or, equivalently,
the covariant conservation law for the momentum current
$P^{\alpha}_x$),
\begin{equation}\label{conservation}
\p_t(\Pi^t_x)+\p_r(\Pi^r_x)=0~,
\end{equation}
 where the first term vanishes due to the trivial time-dependence of
 the embedding
functions (\ref{x}) and (\ref{y}) (while the right-hand side
vanishes because $\cL$ is independent of $X$). {}From this it
becomes clear that the conclusion of the previous paragraph can be
avoided, and the string can lean backward, if and only if it
displays a more complicated time dependence (e.g., some type of
oscillatory behavior).

The basic issue here is a familiar one: specifying boundary
conditions for the string does not select a unique solution to the
corresponding equation of motion; one must additionally specify
initial conditions. If the string is initially upright and moving
as a whole at velocity $v$, then as we have seen it will continue
to do so, and its endpoints will trace the required paths without
requiring an external agent that pulls on them in the direction of
motion. On the other hand, if the string is initially static and
\emph{then} we start pulling its endpoints with whatever force is
necessary for them to move at constant velocity $v$, the string
\emph{will} of course lean backward, as in Fig.~\ref{string}a.
What we have learned above, however, is that such a string will
continue to oscillate and will never (classically) stabilize to a
configuration of the type (\ref{x}). The solution in question is
therefore clearly not the one with the lowest energy for the given
boundary conditions. Nevertheless, as we will see in the next
section, there are circumstances under which this type of solution
might conceivably play a role in the computation of the energy for
the quark-antiquark system of interest to us.

We should similarly keep in mind that it is possible to satisfy
the boundary conditions (\ref{bc}) with two \emph{separate}
strings that reach all the way down from the boundary to the
horizon at fixed $Y=\mp L/2$, trailing behind their endpoints as
described in \cite{hkkky,gubser}. Such configuration would clearly
describe an unbound quark and antiquark, and it will also be of
relevance below.

Let us now proceed towards the determination of the shape of the
moving string in the AdS-Schwarzschild background. Using the
explicit form of the metric\footnote{The relevant background is as
usual (AdS-Schwarzschild)$_5\times \bS^5$, but we display only the
metric for the first factor because the string is taken to lie at
a fixed position on the $\bS^5$.}
\begin{eqnarray}\label{metric}
ds^2&=&{1\over\sqrt{H}}\left(
-hdt^2+d\vec{x}^2\right)+{\sqrt{H}\over h}dr^2~, \\
H&=&{R^4\over r^4}~,\qquad  h=1-\frac{r_H^{4}}{r^4}~, \nonumber
\end{eqnarray}
together with the embedding functions (\ref{x}) and (\ref{y}), the
Lagrangian density (\ref{ng}) simplifies to
\begin{equation}\label{lagrangianfull}
\cL=-\sqrt{-g}=-\sqrt{1+\frac{h}{H}\left(X^{'2}+Y^{'2}\right)
-\frac{v^2}{H}Y^{'2}-\frac{v^2}{h}}~.
\end{equation}
The associated non-vanishing canonical momentum densities are
\begin{eqnarray} \label{pisfull}
\Pi^{t}_{t}&=&{-1\over\sqrt{-g}}\left[1+\frac{h}{H}\left(X^{'2}+Y^{'2}\right)
\right]~, \nonumber\\
\Pi^{t}_{x}&=&{v\over\sqrt{-g}}\left[{1\over h} +\frac{1}{H}Y^{'2}
\right]~, \nonumber\\
\Pi^{r}_{t}&=&{v\over\sqrt{-g}}\left[\frac{h}{H}X^{'}\right]~,\\
\Pi^{r}_{x}&=&{-1\over\sqrt{-g}}\left[\frac{h}{H}X^{'}\right]~,\nonumber\\
\Pi^{r}_{y}&=&{-1\over\sqrt{-g}}\left[\frac{h-v^2}{H}Y^{'}\right]~.\nonumber
\end{eqnarray}

As already noted above, given that the string moves at constant
velocity and $\cL$ is independent of $X$, the corresponding
equation of motion (\ref{conservation}) is just the statement that
the conjugate momentum density $\Pi^r_x$ is a (real) constant,
which we will denote $\Pi_x$ in what follows. The same is of
course true for $\Pi_y\equiv\Pi^r_y$. According to (\ref{forceh}),
these constants determine the forces $-F_x$ and $-F_y$ that an
external agent must exert on the string endpoints to satisfy the
given Dirichlet boundary conditions (\ref{bc}). This agent
supplies energy to the string at a rate $dE/dt=\Pi^r_t/2\pi\ap$,
which as expected is seen from (\ref{pisfull}) to equal the work
$-vF_x$.

Inverting the relations (\ref{pisfull}) to express $X'$ and $Y'$
in terms of the constants $\Pi_x$ and $\Pi_y$, we obtain
\begin{equation}\label{ecuaciondex}
X'=-\Pi_x \frac{(h-v^2)}{h}\sqrt{
\frac{H}{(h-v^2)(\frac{h}{H}-\Pi_x^2)-h\Pi_y^2}}~,
\end{equation}
\begin{equation}\label{ecuaciondey}
Y'=-\Pi_y\sqrt{ \frac{H}{(h-v^2)(\frac{h}{H}-\Pi_x^2)-h\Pi_y^2}}~,
\end{equation}
where we notice the appearance of the same factor $h-v^2$ whose
vanishing defined the critical radius (\ref{rv}) that as explained
above played a crucial role in fixing the value
\begin{equation} \label{gub}
\Pi_x=-\frac{v}{\sqrt{1-v^2}}\left({r_H\over R}\right)^2
\end{equation}
for the string dual to a moving quark \cite{hkkky,gubser}.

There are two immediate things we can learn from the above
equations. First, we see from (\ref{ecuaciondey}) that $Y'$
diverges at the point where the denominator vanishes; according to
the characterization following (\ref{y}) this defines the
turnaround point $r_{min}$:
\begin{equation} \label{rminfull}
\left[(h-v^2)\left(\frac{1}{H}-{\Pi_x^2\over
h}\right)\right]_{r=r_{min}}=\Pi_y^2~.
\end{equation}
It is easy to show from this expression that $r_{min}\ge r_v$,
where equality holds only if $\Pi_y=0$ (which would imply $L=0$).
So, as expected, we find that the string dual to the moving
quark-antiquark pair cannot penetrate beyond the critical radius
$r_v$.

Second, in order for the projections of the two halves of the
string onto the $y-x$ plane to join smoothly, we must require that
$\p Y/\p X=Y'/X'=\infty$ at $r_{min}$, but taking the quotient of
(\ref{ecuaciondey}) and (\ref{ecuaciondex}) we see that this is
\emph{not} possible unless $\Pi_x=0$, and therefore $X'=0$. So, as
we had already anticipated, we learn that the string can only move
at constant speed if it is upright. This same conclusion has been
reached independently in \cite{sonnenschein}.

Specializing (\ref{ecuaciondey}), (\ref{rminfull}),
(\ref{lagrangianfull}) and (\ref{pisfull}) to the case $X'=0$
($\Rightarrow \xi(r)=0$), we can now derive the equations that
will be of interest to us in the remainder of this paper. The
profile of the upright $\cap$-shaped string that moves with
velocity $v$ and has endpoint separation $L$ is determined by
\begin{equation} \label{yprime}
Y'=-\Pi_y\frac{R^4}{\sqrt{(r^4-r^4_H)(r^4(1-v^2)-r^4_H-R^4\Pi^2_y)}}~,
\end{equation}
where the value of $\Pi_y$ must be chosen in such a way that
\begin{equation} \label{L}
L=2\int_{r_{min}}^{\infty}dr\,Y'=2\Pi_y
R^4\int_{r_{min}}^{\infty}\frac{dr}
{\sqrt{(r^4-r^4_H)(r^4(1-v^2)-r^4_H-R^4\Pi^2_y)}}~,
\end{equation}
with
\begin{equation} \label{rmin}
r_{min}=\left(r^4_H+R^4\Pi^2_y\over 1-v^2\right)^{1/4}~.
\end{equation}

Using (\ref{yprime}), the Lagrangian density
(\ref{lagrangianfull}) reduces to
\begin{equation} \label{Lbound}
\cL_{bound}=-\frac{r^4(1-v^2)-r^4_H}{\sqrt{(r^4-r^4_H)
(r^4(1-v^2)-r^4_H-R^4\Pi^2_y)}}~.
\end{equation}
Close to the boundary we find $\cL_{bound}\to-\sqrt{1-v^2}$, which
upon integration implies that the total worldsheet area per unit
boundary time is linearly divergent--- an obvious consequence of
the fact that the string extends all the way to spatial infinity.
The same divergence is found in the area of the two disconnected
worldsheets dual to the unbound quark and antiquark, described by
(\ref{lagrangianfull})-(\ref{ecuaciondey}) with $\Pi_y=0$ and
$\Pi_x$ as in (\ref{gub}), which result in
$\cL_{unbound}=-\sqrt{1-v^2}$. Subtracting the two areas
 we find the finite expression\footnote{In more
 accurate language, one should as usual introduce
a regulating surface at a large radius $r=r_{\Lambda}$ to make
both integrals finite, subtract, and in the end take
$r_{\Lambda}\to\infty$. In the dual gauge theory, this is
equivalent to introducing a UV cutoff $\Lambda\simeq
r_{\Lambda}/R^2$.}
\begin{equation} \label{wilson}
A=-{2\over 2\pi\ap}\int_{-\cT/2}^{+\cT/2}dt\left(
\int_{r_{min}}^{\infty}dr\,\cL_{bound}-\int_{r_{H}}^{\infty}dr\,
\cL_{unbound}\right)~.
\end{equation}
According to the standard recipe \cite{maldawilson}, in the dual
finite-temperature gauge theory, the relative area (\ref{wilson})
determines the expectation value (in a stationary phase
approximation) of the Wilson loop traced by the moving
quark-antiquark pair.

We are also interested in computing the total energy of the
$\cap$-shaped string, which translates into the energy of the
quark-antiquark pair. Starting from (\ref{pisfull}), the
Hamiltonian density $\cH\equiv -\Pi^t_t$ works out to
\begin{equation} \label{Hbound}
\cH_{bound}=\frac{r^4-r^4_H}{\sqrt{
r^4(1-v^2)-r^4_H-R^4\Pi^2_y}}~.
\end{equation}
As expected, the behavior of this expression near the boundary
 gives rise to a linear divergence in the total energy of the
string, which could again be cancelled by subtracting the energy
 of the disconnected strings dual to the unbound quark and
 antiquark, obtained from integrating
\begin{equation} \label{Hunbound}
\cH_{unbound}=\frac{r^4-r^4_H(1-v^2)}{(r^4-r^4_H)\sqrt{1-v^2}}~.
\end{equation}
This subtraction, however, would introduce a new infinity, because
(\ref{Hunbound}) implies that the energy of each of the moving
unbound strings is logarithmically divergent at the $r=r_H$
endpoint of the integration \cite{hkkky}. The physical origin of
this divergence is the infinite amount of energy that has been
provided to the system by the external agent that has pulled the
boundary endpoint of the string along the $x$ direction for an
infinite period of time. {}From the perspective of a boundary
observer, over the course of time this energy has flowed along the
string and accumulated in the vicinity of the horizon.

As explained in \cite{hkkky}, a simple estimate of the work done
on the trailing string by the external agent is obtained by
assuming that it has exerted precisely the force needed to
overcome the constant drag force (\ref{gub}) over exactly the
(infinite) distance that separates the front (boundary) and back
(horizon) endpoints of the string. A short calculation shows that
this amounts to identifying
\begin{equation} \label{Hinput}
\cH^{input}_{unbound}=\frac{r^4_H v^2}{(r^4-r^4_H)\sqrt{1-v^2}}
\end{equation}
as the energy density provided by the external agent. Subtracting
this from (\ref{Hunbound}), we obtain an estimate of the energy
density `intrinsic' to the moving string,
\begin{equation} \label{Hintrinsic}
\cH^{intrinsic}_{unbound}\equiv \cH_{unbound}-
\cH^{input}_{unbound}={1\over\sqrt{1-v^2}}~,
\end{equation}
which, as expected, no longer includes the
logarithmically-divergent portion. The prescription for
eliminating this divergence is of course highly non-unique: one
may add to (\ref{Hinput}) \emph{any} function $\cU(r,v)$ such that
$U(v)\equiv\int_{r_H}^{\infty} \cU(r,v)<\infty$ (in order for
$\cU$ to represent a finite renormalization of the string energy)
and $\cU(r,0)=0$ (to continue to match the known energy of the
static string).

For use below, it is convenient to note here that a completely
analogous story applies to the linear momentum $\cP\equiv\Pi^t_x$
of the unbound strings \cite{hkkky}: upon integration, the
momentum density $\cP_{unbound}=v/(h\sqrt{1-v^2})$ which follows
from (\ref{pisfull}) leads to both a linear divergence at
$r\to\infty$ and a log divergence at $r=r_H$; the latter is a
reflection of the infinite amount of momentum provided by the
external agent, which can be estimated to be
$\cP^{input}_{unbound}=\cH^{input}_{unbound}/v$; the remaining
$\cP^{intrinsic}_{unbound}=v/\sqrt{1-v^2}$ is then an estimate of
the momentum density intrinsic to the moving string.

The preceding discussion points towards
\begin{equation} \label{Eplasma}
E\equiv {2\over 2\pi\ap}\left(
\int_{r_{min}}^{\infty}dr\,\cH_{bound}-\int_{r_{H}}^{\infty}dr\,
\cH^{intrinsic}_{unbound}+U(v)\right)
\end{equation}
as a simple and finite expression that codifies the energy of the
moving $\cap$-shaped string relative to that of the two moving
disconnected strings, or, in dual language, the energy of the
quark-antiquark pair relative to that of the unbound quark and
antiquark. This definition captures, for any given $v$, the
correct $L$-dependence of the energy of the bound system. The
arbitrariness involved in the choice of the function $U(v)$ leads,
however, to two important drawbacks: it denies meaning to a direct
comparison of  values of the energy computed at different
velocities, and makes it impossible to deduce from the value of
$E(L,v)$ alone whether, for a given $L$ and $v$, the energetics
favor the bound or the unbound configuration. The resolution of
these problems will require establishing an unequivocal
operational definition of the intrinsic energy of the moving
unbound strings (a natural suggestion was made in \cite{hkkky}).

As we have seen, the source of the ambiguity in the definition of
$E(L,v)$ is the infinite amount of energy supplied to the unbound
strings by the agent that drags them, so a natural way to sidestep
this difficulty is to compute the energy in the string rest frame,
where the external agent does no work. The requisite coordinate
transformation is of course
\begin{eqnarray}\label{boost}
\bar{t}&=&\gamma(t-v x)~,\nonumber\\
\bar{x}&=&\gamma(x-v t)~,\\
\bar{y}&=&y~,\nonumber\\
\bar{r}&=&r~, \nonumber
\end{eqnarray}
(with $\gamma\equiv 1/\sqrt{1-v^2}$) and amounts, from the gauge
theory point of view, to a Lorentz boost that takes us from the
rest frame of the plasma, where we had worked up to now, to the
rest frame of the quark and antiquark. The canonical momentum
densities (\ref{pisfull}) transform according to
\begin{eqnarray} \label{piboost}
\bar{\Pi}^{\bar{\tau}}_{\bar{\mu}}&=&{\p
X^{\nu}\over\p\bar{X}^{\bar{\mu}}}\left({\p\sigma\over\p\bar{\sigma}}
\Pi^{\tau}_{\nu}-{\p\tau\over\p\bar{\sigma}}
\Pi^{\sigma}_{\nu}\right)~,\\
\bar{\Pi}^{\bar{\sigma}}_{\bar{\mu}}&=&{\p
X^{\nu}\over\p\bar{X}^{\bar{\mu}}}\left({\p\tau\over\p\bar{\tau}}
\Pi^{\sigma}_{\nu}-{\p\sigma\over\p\bar{\tau}}
\Pi^{\tau}_{\nu}\right)~, \nonumber
\end{eqnarray}
where we have taken into account the effect of the change from the
static gauge in the plasma rest frame ($\tau=X^0,\sigma=R$) to the
static gauge in the quark rest frame
($\bar{\tau}=\bar{X}^0,\bar{\sigma}=\bar{R}$).

It is easy to check that $\bar{\Pi}^{\bar{r}}_{\bar{t}}=0$, which
shows that, as expected, in this frame no energy is being supplied
to the string. The total energy of the unbound strings will
consequently have no logarithmic divergence, and its linear
divergence will serve to cancel that of the $\cap$-shaped string
in the usual straightforward way. Because the string is static, we
find $\bcH=-\bar{\Pi}^{\bar{t}}_{\bar{t}}=-\bcL$, and from the
fact that the Lagrangian transforms as a scalar density it follows
that $\bcL=\gamma\cL$, so the energy of the bound system relative
to that of the unbound system,
\begin{equation} \label{ebar}
\bar{E}\equiv-{2\over 2\pi\ap}\left(
\int_{r_{min}}^{\infty}dr\,\bcL_{bound}-\int_{r_{H}}^{\infty}dr\,
\bcL_{unbound}\right)~,
\end{equation}
is related to the area (\ref{wilson}) through $\bar{E}=\gamma
A/\cT=A/\bar{\mathcal{T}}$, just like it should. We stress that in
this frame we have been able to cleanly subtract the energy of the
unbound strings without introducing any ambiguities, so both the
$L$- and $v$-dependence of (\ref{ebar}) are physically meaningful,
and the bound configuration is known to be energetically preferred
whenever $\bar{E}(L,v)<0$.

It is interesting to note that, in contrast with the energy, even
in the rest frame the linear momentum of the string \emph{cannot}
be defined unequivocally without additional physical input. Even
though the external agent does no work on either the bound or
unbound strings, in the latter case it does supply momentum to the
static string: in gauge theory language, a force must be exerted
to hold the isolated quark in place as the plasma flows by at
speed $v$. As a result, the momentum density
$\bcP_{unbound}=v\gamma^2(1-h)/h$ for the disconnected strings
gives rise to a logarithmic divergence at the $r=r_H$ endpoint of
the integration. This may be eliminated by subtracting the
estimate
\begin{equation} \label{Pinput}
\bcP^{input}_{unbound}=v\gamma^2\left({1-h\over h}\right)
\end{equation}
for the momentum supplied by the external agent, which can be
obtained either by Lorentz-transforming
$(\Pi^{\alpha}_{\mu})^{input}_{unbound}$ to the barred frame, or
by recomputing directly in the barred frame under assumptions
parallel to those that led to (\ref{Hinput}). After this
subtraction, one would be left with $
\bcP^{intrinsic}_{unbound}=0$ as an estimate of the momentum
intrinsically associated with the string.  This vanishing result
might at first sight appear natural and unambiguous, since the
string is, after all, at rest. That the issue is not this simple
becomes clear upon observing that the momentum density for the
$\cap$-shaped string,
\begin{equation} \label{Pbound}
\bcP_{bound}=v\gamma(1-h)\cH,
\end{equation}
is non-vanishing, despite the fact that this string is also at
rest, and no external momentum has been supplied to it. This is
only possible because in the barred frame $\bar{g}_{\bt\bx}\neq
0$, so the metric is not static. We will come back to this
discussion in the next section.

\section{Energy of Moving Quark-Antiquark System}
\label{qqbarsec}

In this section we will use the above results for the moving
string to make various inferences about the dual system: an
external quark-antiquark pair in $SU(N)$ $\cN=4$  SYM with
`t~Hooft coupling $g_{YM}^2 N$ and temperature $T$ determined by
the AdS radius $R$, horizon radius $r_H$ and  string length
$\sqrt{\ap}$ through \cite{magoo}
\begin{equation} \label{dict}
g_{YM}^2 N = {R^4\over\ap^2}~,\qquad T={r_H \over\pi R^2}~.
\end{equation}

 {}According to (\ref{forceh}), the fact
that $\Pi_x=0$ translates into the statement that, in stark
contrast with the solitary quark considered in
\cite{hkkky,gubser}, an external quark-antiquark pair feels no
drag force as it ploughs through the plasma, a curious result that
was obtained independently in the recent paper
\cite{sonnenschein}.

As explained in \cite{hkkky,gubser} and analyzed more closely in
\cite{gubseretal1,gxz,gubseretal2}, a moving quark produces an
extended wake in the color fields, which may be regarded as a
coherent spray of gluons radiated away from the quark, and is the
CFT dual of the trailing string that extends all the way down to
the horizon.  It is this wake that transports energy and momentum
away from the quark and into the surrounding medium. It is worth
noting that this mechanism can operate even at zero temperature,
where there is no plasma: given appropriate initial conditions for
the gluonic fields, a quark moving at constant speed \emph{can}
lose energy to the wake it imprints on these fields. The dual
statement is that a string on pure AdS that displays a non-trivial
time-dependence \emph{can} trail behind its boundary endpoint even
if the latter moves at constant velocity.\footnote{The assertion
for the string may be deduced from an argument similar to the
discussion following (\ref{conservation}); its SYM dual could be
verified through calculations similar to those performed in
\cite{cg}.} This, of course, should not come as a surprise,
because the SYM vacuum constitutes, after all, a highly nonlinear
polarizable medium. Needless to say, the lowest-energy
configuration for the gluonic field profile surrounding the moving
quark at zero temperature is the one obtained by boosting the
static profile; this configuration is dual to an upright string,
which feels no drag force. The string considered in
\cite{hkkky,gubser} correctly reduces to this case when $T\to 0$,
which ensures that the energy loss process studied there is
intrinsically associated with the presence of the plasma.

Unlike the single quark, which carries a net color charge, the
quark-antiquark pair is a dipole, and consequently sets up a
shorter-ranged profile in the gluonic fields. At zero-temperature,
the dipolar $\tr F^2$ falloff is proportional to $L^3/r^{7}$
\cite{cg,kmt}, compared to the Coulomb-like $1/r^{4}$ of the
monopole \cite{dkk}. At finite temperature, we have learned here
that the profile generated by the moving pair is not able to
transport energy away from it, a property that could plausibly be
verified using the methods of \cite{gubseretal1,gxz,gubseretal2}.
In the $N\gg 1$, $g_{YM}^2 N\gg 1$ regime of the gauge theory that
is captured by classical string theory on weakly-curved
AdS-Schwarzschild, no other mechanism of energy loss is at work,
and so the quark-antiquark pair moves through the plasma
unimpeded. This result should generalize to any color-neutral
probe of the plasma, including the baryon, whose static AdS dual
was constructed in \cite{wittenbaryon,imamura,cgs} and whose
zero-temperature $\tr F^2$ falloff is also $\propto r^{-7}$
\cite{cg}. The remarks we made above for the solitary quark at
$T=0$ apply as well to these color-neutral systems at $T>0$: with
a different set of initial conditions for the gluonic fields, the
quark-antiquark pair and the baryon \emph{can} experience a drag
force.

Let us now proceed to determine the energies $\bar{E}$ and $E$ of
the quark-antiquark pair for a given velocity $v$ and separation
$L$. For this we first need to carry out the integrals (\ref{L}),
(\ref{ebar}) and (\ref{Eplasma}) to find $L(\Pi_y,v)$,
$\bar{E}(\Pi_y,v)$ and $E(\Pi_y,v)$, and then eliminate $\Pi_y$ to
obtain $\bar{E}(L,v)$ and $E(L,v)$. As indicated in (\ref{forceg})
and explained in the discussion below (\ref{pisfull}), $\Pi_y$ is
a measure of the force $F_y$ that an external agent must exert in
order to keep the $q\bar{q}$ pair at the desired separation.

Unfortunately, the integrals cannot be performed analytically, so
we must solve the problem numerically. For this purpose, it is
convenient to use $h=1-r_H^4/r^4$ in place of $r$ as the
integration variable. The range of integration should then be
taken from
\begin{equation} \label{hmin}
h_{min}\equiv h(r_{min})=\frac{v^2+f_y^2}{1+f_y^2}
\end{equation}
to 1, where we have defined the rescaled force
$$f_y\equiv{R^2\over r_H^2}\Pi_y= {2\over \pi \sqrt{g_{YM}^2 N} T^2}F_y~.$$
The result $r_{min}> r_v$ of the previous section translates into
$h_{min}> v^2$, which can also be easily deduced from
(\ref{hmin}). After changing variables in this manner and using
the dictionary (\ref{dict}), the expression for the
quark-antiquark separation (\ref{L}) turns into
\begin{equation}\label{longitud}
L(f_y,v)=\frac{f_y}{2{\pi}T}\int^{1}_{h_{min}}
{\frac{dh}{(1-h)^{\frac{1}{4}}\sqrt{(h-v^2)h-(1-h)h{f_y^2}}}}~,
\end{equation}
and the energy of the $q\bar{q}$ pair in its rest frame
(\ref{ebar}) becomes\footnote{An overall factor of $\gamma$ was
missing from the energies computed in the first version of this
paper that was posted on the arXiv. We thank Hong Liu for bringing
this to our attention.}
\begin{equation}\label{energia}
\bar{E}(f_y,v)=\frac{T\sqrt{g^2_{YM}N}}{4}\left[
\int^{1}_{h_{min}}{\frac{dh(h-v^2)\gamma}{(1-h)^{\frac{5}{4}}
\sqrt{(h-v^2)h-(1-h)h{f_y^2}}}}
-\int^{1}_{0}{\frac{dh}{(1-h)^{\frac{5}{4}}}}\right]~.
\end{equation}
As noted at the end of the previous section, the subtraction
implemented by the second term in equation (\ref{energia}) ensures
a finite result and corresponds to removing the self-energies of
the quark and antiquark separately held in place as the plasma
flows by with velocity $v$ in the $-x$ direction. The energy of
the system in the frame where the plasma is static and the pair
moves is given instead by (\ref{Eplasma}), which translates into
\begin{equation}\label{energia2}
E(f_y,v)-U(v)=\frac{T\sqrt{g^2_{YM}N}}{4}\left[
\int^{1}_{h_{min}}{\frac{dh\sqrt{h}}{(1-h)^{\frac{5}{4}}
\sqrt{(h-v^2)-(1-h){f_y^2}}}} -\int^{1}_{0}{\frac{
dh\gamma}{(1-h)^{\frac{5}{4}}}}\right].
\end{equation}
As explained below (\ref{Hintrinsic}), the function $U(v)$
reflects an ambiguity in separating the energy intrinsically
associated with the moving quark from the energy supplied by the
agent that pulls the quark and lost to the plasma.

Notice that while the second terms in the plasma frame energy
(\ref{energia2}) and the $q\bar{q}$ frame energy (\ref{energia})
are proportional to one another, the first terms are not. The
reason is that the boost that takes us back from the $q\bar{q}$
frame to the plasma frame mixes the energy
$\int_{h_{min}}^{1}dh\,\bcH_{bound}$ of the pair with its momentum
$\int_{h_{min}}^{1}dh\,\bcP_{bound}$, which as seen in
(\ref{Pbound}) is non-vanishing. Since by definition the quark and
antiquark are at rest in this frame, it is clear that the momentum
in question is carried not directly by them, but by the gluonic
field configuration produced by their interaction with the flowing
plasma, i.e., the momentum density $\bcP(r)$ encodes the
chromodynamic analog of the electromagnetic Poynting vector, at
the energy scale $\sim r/R^2$.\footnote{This also leads one to
expect the momentum intrinsic to the isolated quark held fixed in
the flowing plasma to be non-vanishing, unlike what the naive
estimate (\ref{Pinput}) (which corresponds to $U(v)=0$) would have
indicated.} It would be interesting to explore this relation in
more detail using the methods of \cite{gubseretal2}.

Even though (\ref{energia}) and (\ref{energia2}) are not
proportional to one another, they turn out to be related through
the relatively simple expression
\begin{equation} \label{energias}
E(L,v)-U(v)=\gamma\bar{E}(L,v)+{\pi\over 2}\sqrt{g_{YM}^2
N}T^2{v^2 \gamma^2 L\over f_y}~.
\end{equation}
This enables one to determine $E(L,v)$ once $\bar{E}(L,v)$ is
known, without having to carry out any additional numerical
integration, so in the remainder of this paper we will concentrate
on computing the latter.

\begin{figure}[htb]
\begin{center}
\setlength{\unitlength}{1cm}
\includegraphics[width=6cm,height=4cm]{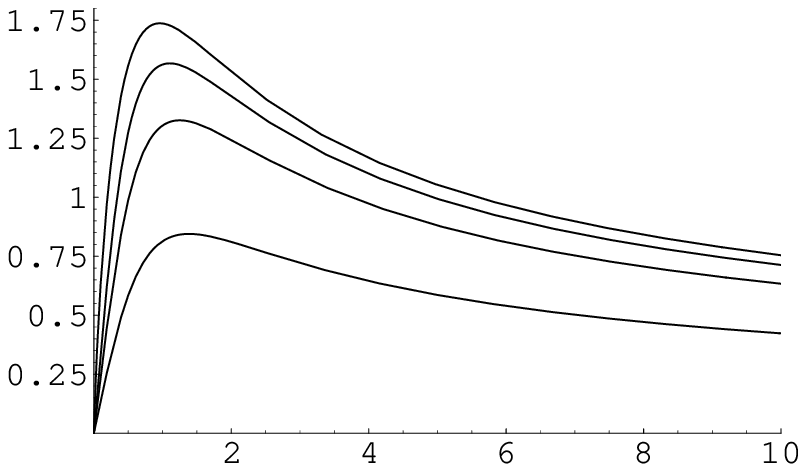}
 \begin{picture}(0,0)
   \put(0.2,0.3){$f_y$}
   \put(-5.5,4.1){$l$}
 \end{picture}
\caption{Quark-antiquark separation (in units of $1/2\pi T$) as a
function of the applied external force (in units of $\pi
\sqrt{g_{YM}^2 N}T^2/2$), for velocities $v=0,0.45,0.7,0.95$.
Lower curves correspond to larger velocities.} \label{lf}
\end{center}
\end{figure}

The results of the numerical integration of (\ref{longitud}) are
shown in Fig.~\ref{lf}, which displays
$$l\equiv 2\pi T L$$
 as a
function of the applied external force $f_y$ for a few different
values of $v$. The behavior is in all cases qualitatively the same
as was found in \cite{theisen,brandhuber} for the static case: at
any given $v$, it is only possible to attain separations in a
finite range $0\le L\le L_{max}(v)$, and each separation in this
range can be achieved with two different values of the force
$F_y$. The exception is of course the maximum $L_{max}(v)$, whose
physical meaning will become clear below, and which we find
empirically to be located at a value of the external force that
can be well-approximated with a quadratic function of the
velocity,
\begin{equation} \label{fmax}
{f_y}_{max}(v)\simeq 0.949+0.247v+0.223v^2~.
\end{equation}

\begin{figure}[htb]
\setlength{\unitlength}{1cm}
\includegraphics[width=6cm,height=4cm]{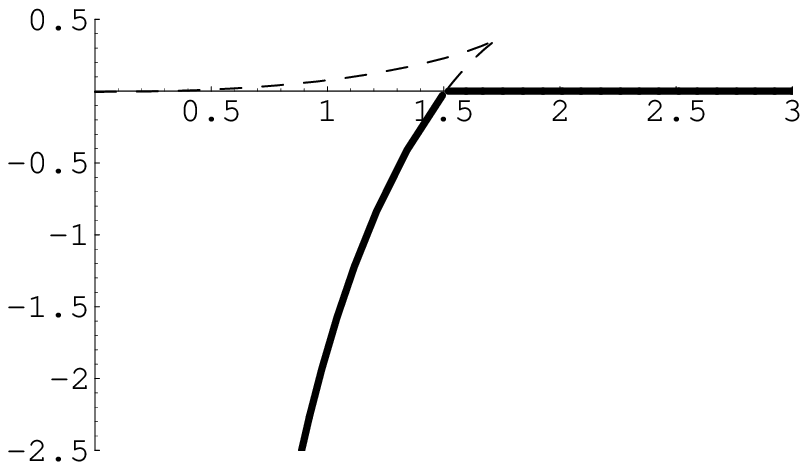}
 \begin{picture}(0,0)
   \put(0.2,3){$l$}
   \put(-5.5,4){$\bar{e}$}
 \end{picture}\hspace{1cm}
 \includegraphics[width=6cm,height=4cm]{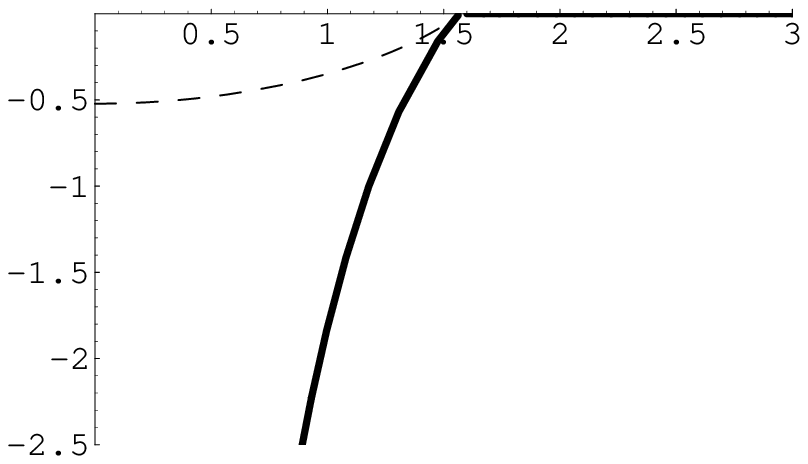}
 \begin{picture}(0,0)
   \put(0.2,3.6){$l$}
   \put(-5.5,4){$\bar{e}$}
 \end{picture}
\caption{Quark-antiquark energy (in units of $T\sqrt{g_{YM}^2
N}/4$) as a function of separation (in units of $1/2\pi T$), for
(a) $v=0$ (b) $v=0.45$. The solid (dashed) portion of each curve
corresponds to stable (metastable) configurations.} \label{elow}
\end{figure}

\begin{figure}[htb]
\setlength{\unitlength}{1cm}
\includegraphics[width=6cm,height=4cm]{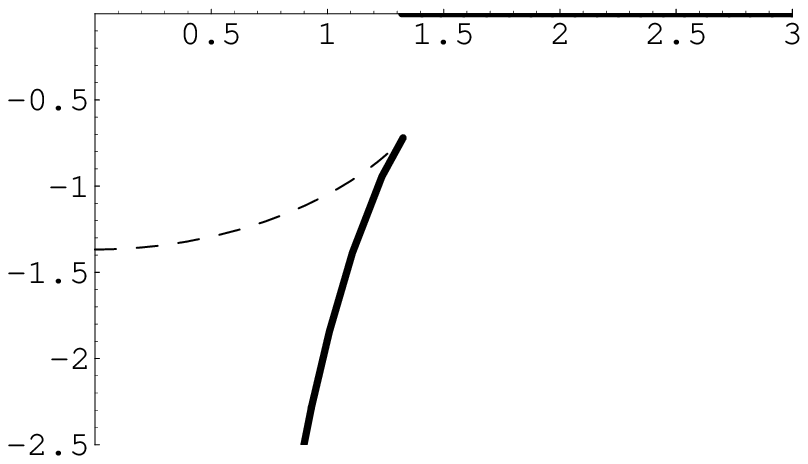}
 \begin{picture}(0,0)
   \put(0.2,3.6){$l$}
   \put(-5.5,4){$\bar{e}$}
 \end{picture}\hspace{1cm}
 \includegraphics[width=6cm,height=4cm]{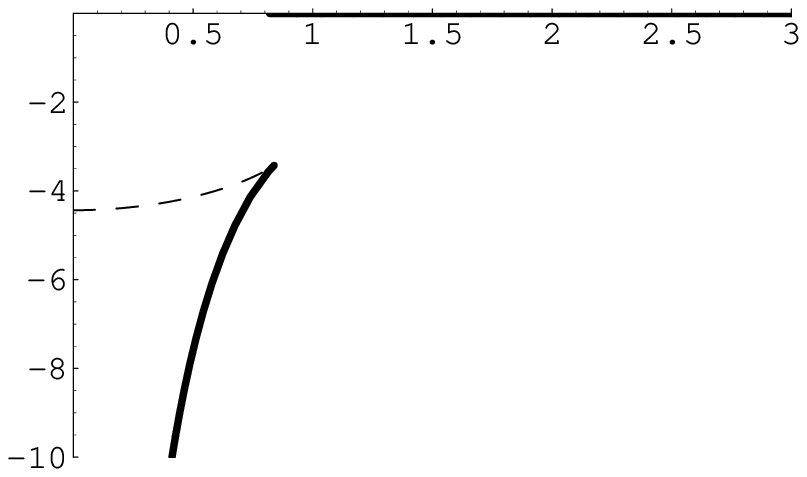}
 \begin{picture}(0,0)
   \put(0.2,3.6){$l$}
   \put(-5.5,4){$\bar{e}$}
 \end{picture}
\caption{Quark-antiquark energy (in units of $T\sqrt{g_{YM}^2
N}/4$) as a function of separation (in units of $1/2\pi T$), for
(a) $v=0.7$, (b) $v=0.95$. The solid (dashed) portion of each
curve corresponds to stable (metastable) configurations.}
\label{ehigh}
\end{figure}

Combining these results with the  numerical integration of
(\ref{energia}), we can find the quark-antiquark energy
$\bar{E}(L,v)$ for any velocity $0\le v\le 1$ and separation $0\le
L\le L_{max}(v)$. The results are plotted in Figs.~\ref{elow} and
\ref{ehigh}, which display
$$\bar{e}\equiv{4\over\sqrt{g_{YM}^2 N}T}\bar{E}$$
as a function of $l$, for a few representative values of $v$. In
each case the curve is divided into two parts: a dashed portion
obtained from the smaller value of the applied force consistent
with the given separation $L$ (i.e., $f_y<{f_y}_{max}$), and a
solid portion obtained from the larger value
($f_y\ge{f_y}_{max}$). As we can see in the figures, it is this
latter case that gives the lower value for the quark-antiquark
energy, and consequently the solid curve describes the stable
configurations that are of most interest to us. The dashed curve
is associated instead with configurations that are physical and
can be selected through a proper choice of initial conditions for
the gluonic fields in SYM (or, in dual language, for the string in
AdS-Schwarzschild), but are only metastable (i.e., they are stable
under small, but not arbitrary, fluctuations).

In Fig.~\ref{elow}a we verify that for the static configuration
$v=0$ we correctly reproduce the $q$-$\bar{q}$ potential computed
in \cite{theisen,brandhuber}, which, as explained there, encodes
all of the expected physics. At small separations (large energies)
the quark-antiquark pair becomes insensitive to the plasma and as
a result the potential approaches the $1/L$ behavior obtained at
$T=0$ in \cite{maldawilson}. As the separation grows, however, the
effects of the plasma progressively screen the quark and antiquark
from one another, and as a consequence raise the system's energy
above the Coulombic behavior. The screening becomes complete at
the distance $L_*\approx 1.51/2\pi T <L_{max}(0)$ where the energy
matches that of the unbound system. For separations larger than
this screening length, the quark and antiquark are free and the
$q\bar{q}$ potential vanishes, as indicated by the horizontal
solid line. The configurations described by (\ref{longitud}) and
(\ref{energia}) in the range $L_*<L\le L_{max}$
 are only metastable, which is why the corresponding portion of the
 curve is also dashed.

\begin{figure}[htb]
\begin{center}
\setlength{\unitlength}{1cm}
\includegraphics[width=6cm,height=4cm]{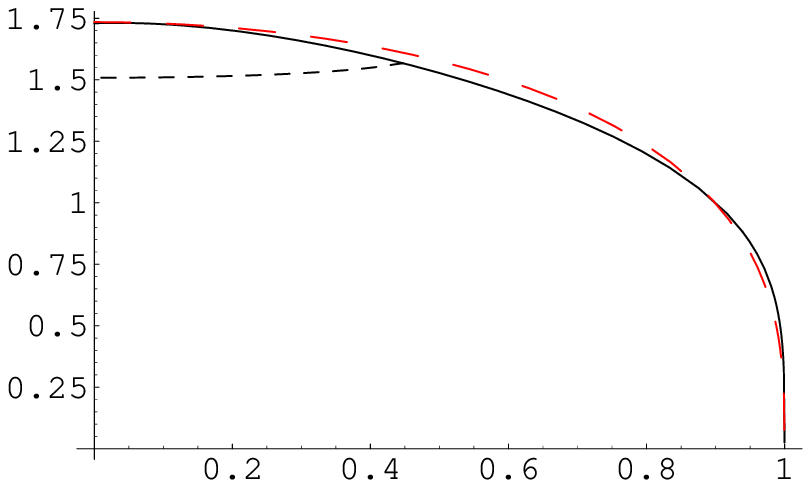}
 \begin{picture}(0,0)
   \put(0.2,0.3){$v$}
   \put(-5. ,4.1){$l_{max}$}
   \put(-5.1,2.9){$l_{*}$}
   \put(-0.4,2.3){$l_{*}=l_{max}$}
 \end{picture}
\caption{Maximum quark-antiquark distance $L_{max}$ and screening
length $L_*$ as functions of the velocity. Both lengths are given
in units of $1/2\pi T$.}\label{lv}
\end{center}
\end{figure}

As seen in Figs.~\ref{elow} and \ref{ehigh}, the results for $v>0$
have many similarities with the static case. The main overall
effect of increasing the velocity is to move the $\bar{E}(L)$
curve to the left and down. The dependence of $L_{max}$ on the
velocity is given by the solid line in Fig.~\ref{lv}. We find it
to be quite close to
\begin{equation} \label{screening}
L_{max}(v)\simeq{1.73\over 2\pi T}(1-v^2)^{1/3}~,
\end{equation}
shown as the long-dash line in the figure. We will comment on the
precise $v\to 1$ behavior below.

\begin{figure}[htbp]
\begin{center}
\setlength{\unitlength}{1cm}
\includegraphics[width=6cm,height=4cm]{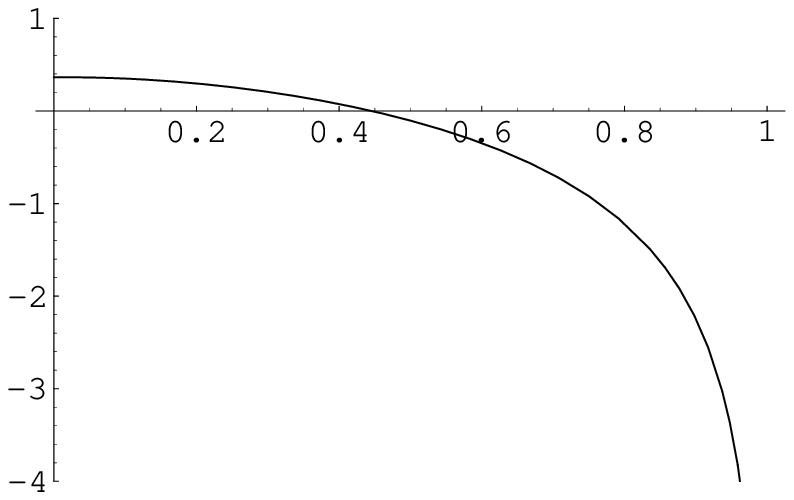}
 \begin{picture}(0,0)
   \put(0.2,2.8){$v$}
   \put(-5.5,4.1){$\bar{e}_{max}$}
 \end{picture}
\caption{Maximum energy as a function of $v$, in units of
$T\sqrt{g^2_{YM}N}/4$. }\label{emax}
\end{center}
\end{figure}

The energy at this maximum separation, $\bar{E}_{max}(v)\equiv
\bar{E}(L_{max},v)$ is shown as a function of velocity in
Fig.~\ref{emax}. For increasing $v$ this energy decreases, passing
through zero at a velocity $\sim 0.447$, and then approaching
$-\infty$ as $v\to 1$. We find that over most of the $0\le v \le
1$ range the graph is practically indistinguishable from that of
the function
\begin{equation} \label{emaxwhole}
\bar{E}_{max}(v)\simeq\frac{0.368T
\sqrt{g^2_{YM}N}}{4}(1-5v^2)(1-v^2)^{-5/12}~.
\end{equation}
The precise behavior in the $v\to 1$ limit will be determined
below.

For any velocity, we expect the presence of the plasma to become
irrelevant at short distances, so at small $L$ our results should
approach the corresponding zero-temperature curves. The latter can
be determined analytically. By taking the $T\to 0$ limit in
 (\ref{longitud}), (\ref{energia}) and (\ref{rmin}) we obtain
\begin{equation}
L=2R^4\Pi_y\int^{\infty}_{r_{min}}\frac{dr}{r^2
\sqrt{(1-v^2)r^4-R^4\Pi^2_y}}
\end{equation}
and
\begin{equation}
\bar{E}=\frac{1}{\pi{\alpha}'}\left[\int^{\infty}_{r_{min}}
{\frac{r^2\sqrt{1-v^2}dr}{\sqrt{(1-v^2)r^4-R^4\Pi^2_y}}}-\int^{\infty}_{0}
dr\right]~,
\end{equation}
with $r_{min}=R\sqrt{\Pi_y}/(1-v^2)^{1/4}$. Changing to the
dimensionless integration variable
$\rho\equiv(1-v^2)^{1/4}r/R\sqrt{\Pi_y}$, it is possible to find
an explicit relation between $\bar{E}$ and $L$,
\begin{equation}\label{Lchica}
\bar{E}(L,v)=-\frac{4{\pi}^2\sqrt{g^2_{YM}N}}
{{\Gamma(\frac{1}{4})}^4L}~,
\end{equation}
which agrees for any $v$ with the static result obtained some
years ago in \cite{maldawilson}. We have checked that our results
for $\bar{E}(L,v)$ at finite temperature correctly approach
(\ref{Lchica}) at small separation. The reason for this agreement
is evident from the string theory side: the limit $L\to 0$ implies
that $r_{min}\to\infty$, so for small separations the string does
not penetrate far into the AdS-Schwarzschild geometry, and it is
difficult for it to sense the difference with pure AdS. Notice
also that, in this limit, the second term in  both the left- and
right-hand side of (\ref{energias}) becomes irrelevant (the
former, because the ambiguity that led to $U(v)$ was associated
with the presence of the horizon; the latter, because it scales
like $L^3$ in this limit), so the plasma frame energy reduces
unequivocally to
\begin{equation}\label{Lchica2}
E(L,v)=\gamma\bar{E}(L,v)=-\frac{4{\pi}^2\sqrt{g^2_{YM}N}}
{{\Gamma(\frac{1}{4})}^4\sqrt{1-v^2}L}~,
\end{equation}
as dictated by the restored Lorentz invariance.\footnote{This
dependence was also noted recently in \cite{sin2}, which includes
some comments on the finite-temperature behavior of $E(L,v)$ in
the ladder approximation of the gauge theory.}

The behavior of the full $\bar{E}(L,v)$ graph for small velocities
is essentially the same as in the static case. As the velocity
increases, the screening length $L_*(v)$ (which we always define
as the separation beyond which the quark and antiquark become
unbound\footnote{Given the shape of the energy graphs, the fact
that $\bar{E}(L_*(v),v)=0$ implies that for separations $L>L_*$
the decay from the bound to the unbound configurations is allowed
from the point of view of energy conservation. Since, as we have
seen above, the momenta of the configurations is in general
non-vanishing (despite the fact that we are in their rest frame),
strictly speaking one would also need to check that the decay is
allowed from the point of view of momentum conservation. This,
however, would require precise knowledge of the momentum intrinsic
to the isolated quark, which is at present lacking. Such knowledge
would also enable one to determine the screening length directly
from the plasma frame energy $E(L,v)$.} is found to increase
slowly, as shown by the short-dash line in Fig.~\ref{lv}. The
dependence is nearly quadratic,
\begin{equation} \label{lstarlow}
L_*(v)\simeq {1.51\over 2\pi T}\left(1+{1\over 3}v^2\right)\quad
\mbox{for $v< 0.447$}.
\end{equation}
As seen in the same figure, $L_{max}(v)$ is monotonically
decreasing, so there is a velocity $v_{gap}\sim 0.447$ at which
both lengths coincide; this is precisely the zero in
Fig.~\ref{emax}, and explains the bound on the region of validity
of (\ref{lstarlow}).

For $v>v_{gap}$, both types of bound solutions (stable and
metastable) have negative energy, so a gap begins to develop
between the bound and unbound configurations, whose width evolves
as indicated in the negative portion of the graph in
Fig.~\ref{emax}. Since the width increases without bound as $v\to
1$, it is natural to wonder whether at $v>v_{gap}$ there could be
additional \emph{bound} $q\bar{q}$ configurations which cover a
range of separations $L>L_{max}$, and consequently narrow or
perhaps even eliminate the gap. As discussed earlier in this
section, there certainly exist configurations in which the quark
and antiquark move at constant velocity but the color fields
display a more complicated time dependence. Their AdS description
was discussed in the paragraphs that follow (\ref{conservation});
it involves a string that leans back as in Fig.~\ref{string}a and
has time dependence beyond the overall motion at velocity $v$.
These configurations exist both for $L\le L_{max}$ and
$L>L_{max}$, but in the former case they are clearly metastable
and therefore not of interest for the present discussion. What is
not at all obvious to us is whether at least one of the
configurations for $L>L_{max}$ manages to have negative energy.
This is a complicated question that would appear to require
numerical exploration of the space of solutions to the
corresponding coupled partial differential equations.

In the remainder of this paper we assume that for all values of
$L$ and $v$, the lowest energy configurations are always the ones
with the simplest time dependence: for $L<L_{max}$, the bound
$q\bar{q}$ system dual to the upright string in
Fig.~\ref{string}b; for $L>L_{max}$,  the unbound quark and
antiquark dual to two separate copies of the string analyzed in
\cite{hkkky,gubser}. The graphs in Figs.~\ref{elow},\ref{ehigh}
can then be taken at face value, and imply in particular that for
$v>v_{gap}$ the screening length should be identified with the
location of the discontinuity in $\bar{E}(L)$, i.e.,
\begin{equation} \label{lstarhigh}
L_*(v)=L_{max}(v)\quad\mbox{for $v>0.447$}.
\end{equation}

The AdS/CFT prediction for the velocity-dependence of the
screening length (for an arbitrary angle $\theta$ between the
direction of motion and the $q$-$\bar{q}$ separation $L$) was the
main subject of \cite{liu2},\footnote{This work also emphasized
the importance of this calculation for advancing towards a
quantitative understanding of the $J/\psi$ suppression observed in
the quark-gluon plasma produced at RHIC.} which appeared while the
first version of this paper was in preparation. The authors of
that work did not compute $\bar{E}(L,v)$, and chose to define the
screening length not as $L_*$ but as the maximum allowed
separation $L_{max}$, throughout the entire range of velocities.
Their equations and numerical results (for $\theta=\pi/2$) are in
complete agreement with ours. As noted in (\ref{screening}), we
have found that, \emph{over the entire range} $0\le v\le 1$, the
$L_{max}(v)$ curve is best described as being proportional to
$(1-v^2)^{1/3}$. The authors of \cite{liu2}, on the other hand,
have found analytically that the behavior of $L_{max}(v)$ \emph{in
the ultra-relativistic limit}\footnote{We thank Hong Liu for
clarifying this point to us after the first version of this paper
was posted on the arXiv.} is precisely proportional to
$(1-v^2)^{1/4}$. This result can be confirmed directly from
(\ref{longitud}), which in the $v\to 1$ limit reduces to
\begin{equation}\label{lhigh}
L(f_y,v)={1\over 2\pi
T}{4\sqrt{2}\pi^{3/2}\over\Gamma(1/4)^2}{f_y\over
(1+f_y^2)^{3/4}}(1-v^2)^{1/4}~.
\end{equation}
In agreement with (\ref{fmax}), this expression has a maximum at
${f_y}_{max}=\sqrt{2}$, which leads to
\begin{equation}\label{lmaxhigh}
L_{max}(v)={1\over 2\pi T}{
3^{-3/4}8\pi^{3/2}\over\Gamma(1/4)^2}(1-v^2)^{1/4}\approx
{1.49\over 2\pi T}(1-v^2)^{1/4}\quad\mbox{for $v\sim 1$.}
\end{equation}
Combining this with (\ref{lstarlow}) and (\ref{lstarhigh}), we
obtain the relatively simple expression
\begin{equation}\label{lstar}
L_*(v)\simeq{1.51\over 2\pi T}\left(1+{7\over 12}v^2-{7\over
12}v^4\right)(1-v^2)^{1/4}\quad\mbox{for $0\le v\le 1$}~,
\end{equation}
which captures the correct analytic behavior at $v\to 0$ and $v\to
1$, and gives good numerical agreement over the entire range of
velocities.

As seen in these last two equations, for large velocities the
screening length $L_*(v)$ decreases monotonically to zero,
implying that $\bar{E}(L,v)=0$ everywhere except in the rapidly
shrinking range $0<L<L_*(v)$, where $\bar{E}(L,v)$ may be obtained
from (\ref{lhigh}) and the $v\to 1$ limit of (\ref{energia}),
\begin{equation} \label{efhigh}
\bar{E}(f_y,v)=-{T\sqrt{g_{YM}^2 N}\over
4}{4\sqrt{2}\pi^{3/2}\over\Gamma(1/4)^2}
{2+f_y^2\over(1+f_y^2)^{3/4}}(1-v^2)^{-1/4}~.
\end{equation}
It is interesting to note that even though $L$ is small and the
condition $r_{min}>r_v=r_H(1-v^2)^{-1/4}$ forces the string to
stay close to the boundary, the $L$-dependence of the energy
remains more complicated than in the $T=0$ case, and is
qualitatively similar to the graphs shown in Fig.~\ref{ehigh}:
there is a metastable region at $f_y<\sqrt{2}$, and a stable
region at $f_y>\sqrt{2}$, such that for $f_y\gg 1$ one recovers
the Coulombic behavior (\ref{Lchica}). By evaluating (\ref{ehigh})
at ${f_y}_{max}=\sqrt{2}$ we find
\begin{equation} \label{emaxhigh}
\bar{E}_{max}(v)=-{T\sqrt{g_{YM}^2 N}\over
4}{2^{9/2}3^{-3/4}\pi^{3/2}\over\Gamma(1/4)^2}(1-v^2)^{-1/4}
\quad\mbox{for $v\sim 1$.}
\end{equation}
Note again the difference in the exponents found here and in
(\ref{emaxwhole}), which was meant as a fit of $\bar{E}_{max}(v)$
throughout the entire interval $0\le v<1$.

In the $v\to 1$ limit the Wilson loop traced by our moving
quark-antiquark pair approaches the lightlike loop used in
\cite{liu} to propose a non-perturbative definition of the
jet-quenching parameter $\hat{q}$.\footnote{To be more precise,
one should take $v\to -1$ to agree with the conventions of
\cite{liu}.} We would have therefore expected our results to make
contact with those of \cite{liu} in this limit, and were surprised
to find that this is not the case. The difference is drastic:
whereas the string of \cite{liu} extends all the way from the
boundary to the horizon and back, in the $v\to 1$ limit the string
that we use to describe bound configurations explores only an
ever-shrinking region of the geometry close to the
boundary.\footnote{One should of course remember that such strings
are allowed only for separations smaller than the screening
length, which approaches zero in the high-velocity limit. For
larger separations the system is unbound, and involves two
separate strings which do extend from the boundary to the horizon,
but are still quite distinct from the string considered in
\cite{liu}.}

The main obstruction to a continuous interpolation between our
results and those of \cite{liu} is the fact that for any value of
$v$ we have found and employed string worldsheets whose area is
real, whereas the authors of \cite{liu} work with a worldsheet
whose area is imaginary. The difference does not appear to be
attributable to the fact that their Wilson loop is strictly
lightlike, while ours is (in general) only approximately so,
because the gauge theory calculations which motivate the
connection to the jet-quenching parameter build upon an eikonal
approximation justified in terms of a high energy limit which
manifestly takes $v\to 1$ from below \cite{kw,urs}. Moreover, an
argument has recently been given in \cite{liu2,krishna} to the
effect that the result for the lightlike Wilson loop of \cite{liu}
can be obtained continuously from Wilson loops that correspond to
velocities that approach $v=1$ from below. The key point is that,
for any given $v< 1$, the authors of \cite{liu,liu2} enforce
boundary conditions for the string not at the AdS-Schwarzschild
boundary, but at a finite radius $r=r_{LRW}\ll r_v$ (with $r_v$
the critical radius given in (\ref{rv})). As a result of this,
their worldsheet lies entirely in the region $r<r_v$, which is
inaccessible to a string that reaches the boundary, as do the
strings considered in the present paper. This explains why the
worldsheets that lead to the result of \cite{liu} are spacelike.

Regrettably, we do not understand the physics behind this
prescription. One can envision of course situations where the
choice of boundary conditions for a path integral result in its
being dominated by a saddle point with imaginary
action,\footnote{A simple example is provided by the computation
of the propagator for a free relativistic particle moving across a
spacelike interval.} but we do not see why this should be the case
in the problem at hand. To determine the value of a Wilson loop
traced by a $q\bar{q}$ pair that moves at any velocity smaller
than, but \emph{arbitrarily close} to, the speed of light, the
AdS/CFT recipe \cite{maldawilson} requires the string boundary
conditions (\ref{bc}) to be enforced at $r\to\infty$, because it
is only in this limit that the dual quark and antiquark become
pointlike. Since this limit is taken at fixed $v$, the string in
question will have no choice but to lie entirely in the $r>r_v$
region, so its worldsheet will be timelike, and the predicted
value for the Wilson loop at strong coupling will unambiguously
coincide with the result $\exp[i\bar{\cT} \bar{E}(L,v)]$ obtained
in this paper.

If the string boundary conditions (\ref{bc}) are enforced instead
as advocated in \cite{liu,liu2,krishna}, the string endpoints lie
at a finite radius $r=r_{LRW} \ll r_v$. In this case the path
integral for the string is indeed dominated by a saddle point with
imaginary action, a condition which has been argued
\cite{liu2,krishna,urs} to be necessary in order to make contact
with the jet-quenching parameter defined in phenomenological
models of energy loss. But by the standard UV/IR reasoning
\cite{uvir}, this path integral would appear to be computing not a
standard but a `thick' Wilson loop, traced by sources for the
gluonic field that have a characteristic size $d\sim
R^2/r_{LRW}\gg R^2/r_{v}\simeq (1-v^2)^{1/4}/T$, which according
to (\ref{lstar}) happens to be much larger than the screening
length at the given $v$. It is unclear to us whether this `thick'
loop is in some way relevant to the approximate gauge theory
calculations \cite{kw} that motivated the proposal of \cite{liu}.
One should not of course lose sight of the fact that the loop
becomes `thinner' (in the sense that $d\to 0$) as $v\to 1$, so
precisely at $v=1$ one is computing a standard Wilson loop (with a
string worldsheet that correctly extends all the way to the
AdS-Schwarzschild boundary). But, as already noted above, the
gauge theory basis for the definition of \cite{liu} would appear
to allow a smooth approach via \emph{standard} Wilson loops with
$v\to 1$ from below, which the AdS/CFT correspondence would
compute using timelike worldsheets up to and including $v=1$
(where we would find $\bar{E}(L,v=1)=0$). Perhaps a more useful
characterization of the $v=1$ Wilson loop computed in \cite{liu}
is as a smooth limit of standard Wilson loops with $v\to 1$ from
\emph{above}.

Before leaving this subject, it is interesting to note that the
$\bar{E}\propto L^2$ dependence that was called for in the
definition of $\hat{q}$ proposed in \cite{liu}--- a dependence
that was successfully obtained in that work using the spacelike
worldsheets discussed in the preceding paragraphs--- can also be
coaxed out of the $v\to 1$ limit of the \emph{timelike}
worldsheets analyzed in this paper, by focusing not on the stable
but on the metastable (dashed) portion of the $\bar{E}(L,v)$
curves of Fig.~\ref{ehigh} that lie near the intersection with the
$\bar{E}$ axis. This region corresponds to configurations with
small separations and small applied external forces, $f_y\ll 1$.
Using this condition it is straightforward to infer from
(\ref{longitud}) and (\ref{energia}) and that, at next-to-leading
order in $L$,
\begin{eqnarray}
\bar{E}(L)&=&\frac{T(g^2_{YM}N)^{\frac{1}{2}}}{4}\left[\int^1_{h_{min}}
{\frac{\gamma\sqrt{h-v^2}dh}{\sqrt{h}(1-h)^{\frac{5}{4}}}}
-\int^1_0{\frac{dh}{(1-h)^{\frac{5}{4}}}}\right]\\
{}&{}&+\frac{L^2T^3{\pi}^2(g^2_{YM}N)^{\frac{1}{2}}}{2}\gamma
\left[\int^1_{h_{min}}{\frac{dh}{(1-h)^{\frac{1}{4}}\sqrt{h(h-v^2)}}}\right]^{-1}~,
\nonumber
\end{eqnarray}
 i.e., the energy depends quadratically on the $q$-$\bar{q}$
 separation, as desired. In the limit $v\to 1$, this relation implies
\begin{eqnarray} \label{q}
\bar{E}(L)&=&{\sqrt{2}\over 4}\left[-(1-v^2)^{-1/4}\mathcal{A}
+(1-v^2)^{-3/4}\mathcal{K}L^2\right]~,\\
\mathcal{A}&=&{8\pi^{3/2}\over\Gamma(1/4)^2}
\sqrt{g^2_{YM}N}T~,\qquad \mathcal{K}={\sqrt{\pi}\over
4}\Gamma(1/4)^2 \sqrt{g^2_{YM}N}T^3~,\nonumber
\end{eqnarray}
where the numerical prefactor in the first equation has been
chosen according to the normalization used in \cite{liu}, in order
to make $\mathcal{K}$ directly comparable to $\hat{q}$.
Independently of whether or not there exists some argument
relating the coefficient $\mathcal{K}$ to the jet-quenching
parameter as defined in phenomenological models \cite{baier,kw},
this calculation shows that the information encoded in the
parameter $\hat{q}$ defined in \cite{liu} can also be accessed
using the timelike worldsheets studied in the present paper. This
is especially interesting in view of the fact that in the $v\to 1$
limit, such worldsheets never wander far from the
AdS-Schwarzschild boundary. Due to the conformal invariance of the
underlying gauge theory, the temperature-dependence of the
parameters $\mathcal{K}$ and $\hat{q}$ was bound to agree. The
agreement in their 't~Hooft-coupling dependence is also not
particularly surprising. What is perhaps worth noting is that the
numerical coefficients are practically equal,
$$\mathcal{K}=(\Gamma(1/4)^4/16\pi^2)\hat{q}\approx 1.1\hat{q}.$$
In the absence of a direct gauge (or string) theory link between
these two parameters, it might
 be worth  exploring their relation in other gauge theories
 with known gravity duals.

\section*{Acknowledgements}

It is a pleasure to thank Elena C\'aceres for collaboration in the
initial stages of this work, for valuable discussions, and for
useful comments on the manuscript. We are also grateful to Chris
Herzog, Hong Liu, Krishna Rajagopal and Urs Wiedemann for helpful
correspondence, and to David Vergara for pointing out a number of
relevant references. This work was partially supported by Mexico's
National Council of Science and Technology (CONACyT) grants
CONACyT 40754-F and CONACyT SEP-2004-C01-47211, as well as by
DGAPA-UNAM grant IN104503-3.

\end{document}